\documentclass[twocolumn, a4paper]{article}

\usepackage{INTERSPEECH2021}
\usepackage{graphicx}
\usepackage{pgfplots}
\usepackage{multirow}

\PassOptionsToPackage{svgnames}{xcolor}
\usepackage{tcolorbox}
\usepackage{lipsum}
\tcbuselibrary{skins,breakable}
\usetikzlibrary{shadings,shadows}

    {\endtcolorbox}
    
\newenvironment{eqblck}[1]{%
    \tcolorbox[
    noparskip,breakable,
    colback=black!5,
    colframe=black!75!black,
    fonttitle=\sffamily\bfseries\normalsize,
    title=#1]}%
    {\endtcolorbox}

\title{AvaTr: One-Shot Speaker Extraction with Transformers}
\name{Shell Xu Hu$^{1*}$\thanks{$^*$Equal contribution.}, Md Rifat Arefin$^{1 2*}$, Viet-Nhat Nguyen$^{1*}$, Alish Dipani$^1$, \\
Xaq Pitkow$^{1 3 4}$ and Andreas Savas Tolias$^{1 4}$}
\address{
  $^1$Upload AI LLC, $^2$University of T\"ubingen,
  $^3$Rice University, $^4$Baylor College of Medicine}
\email{\{shell, rifat.arefin, viet, alish, xaq, andreas\}@uploadai.com}

\begin{document}

\maketitle

\begin{abstract}
  To extract the voice of a target speaker when mixed with a variety of other sounds, such as white and ambient noises or the voices of interfering speakers, 
  we extend the Transformer network \cite{vaswani2017attention} to attend the most relevant information with respect to the target speaker given the characteristics of his or her voices as a form of contextual information. The idea has a natural interpretation in terms of the \emph{selective attention theory} \cite{broadbent1958perception}.
  Specifically, we propose two models to incorporate the voice characteristics in Transformer based on different insights of where the feature selection should take place. Both models yield excellent performance, on par or better than published state-of-the-art models on the \emph{speaker extraction} task, including separating speech of novel speakers not seen during training.
\end{abstract}

\section{Introduction}

A fundamental problem in auditory processing --- the so-called cocktail party effect --- is how to extract target signals under noisy and complex listening conditions, while ignoring interfering noises. 
Example problem formulations include \emph{speech separation} \cite{yu2017permutation,hershey2016deep,luo2019conv}, \emph{speech enhancement} \cite{xu2013experimental,fu2017raw,choi2018phase}, and \emph{speaker extraction} \cite{wang2018voicefilter,xu2020spex,gfeller2020one}. 
In the most popular, speech separation (also known as \emph{blind source separation}), the goal is to decompose the mixed audio into a fixed number of components that generated the sound. This formulation may not be feasible when the number of speakers vary, where the issue is that we do not know which component corresponds to which speaker. 
The speech enhancement problem introduces the concept of signal-to-noise separation, but most existing works assume the noise is non-speech \cite{reddy2020icassp,reddy2021interspeech}. 
 Finally, speaker extraction solves a more realistic signal-to-noise separation problem, utilising pre-recorded clean speech of the target speaker to learn the speaker-specific characteristics which facilitate the extraction of the new speech of that speaker from the mixture of noises and interfering speech from other speakers.


Our paper addresses speaker extraction, with the more difficult constraint that only a snippet of clean speech (e.g., 2 seconds) are supplied for each target speaker. 
The main challenge is how we focus the network on any given target speaker. The most closely related previous work includes VoiceFilter \cite{wang2018voicefilter}, which uses a combination of convolutional and recurrent layers on the Fourier transform of the mixed audio and the voice characteristics extracted from the supplied clean speech; SoundFilter \cite{gfeller2020one} which uses feature-wise linear modulation (FiLM) \cite{perez2018film} for clean speech conditioning; and SpEx \cite{xu2020spex} which inserts the speaker embedding repeatedly into the stack of temporal convolutional modules by concatenating it with the raw features of the mixed audio.

\begin{figure}[t]
  \centering
  \includegraphics[width=\linewidth]{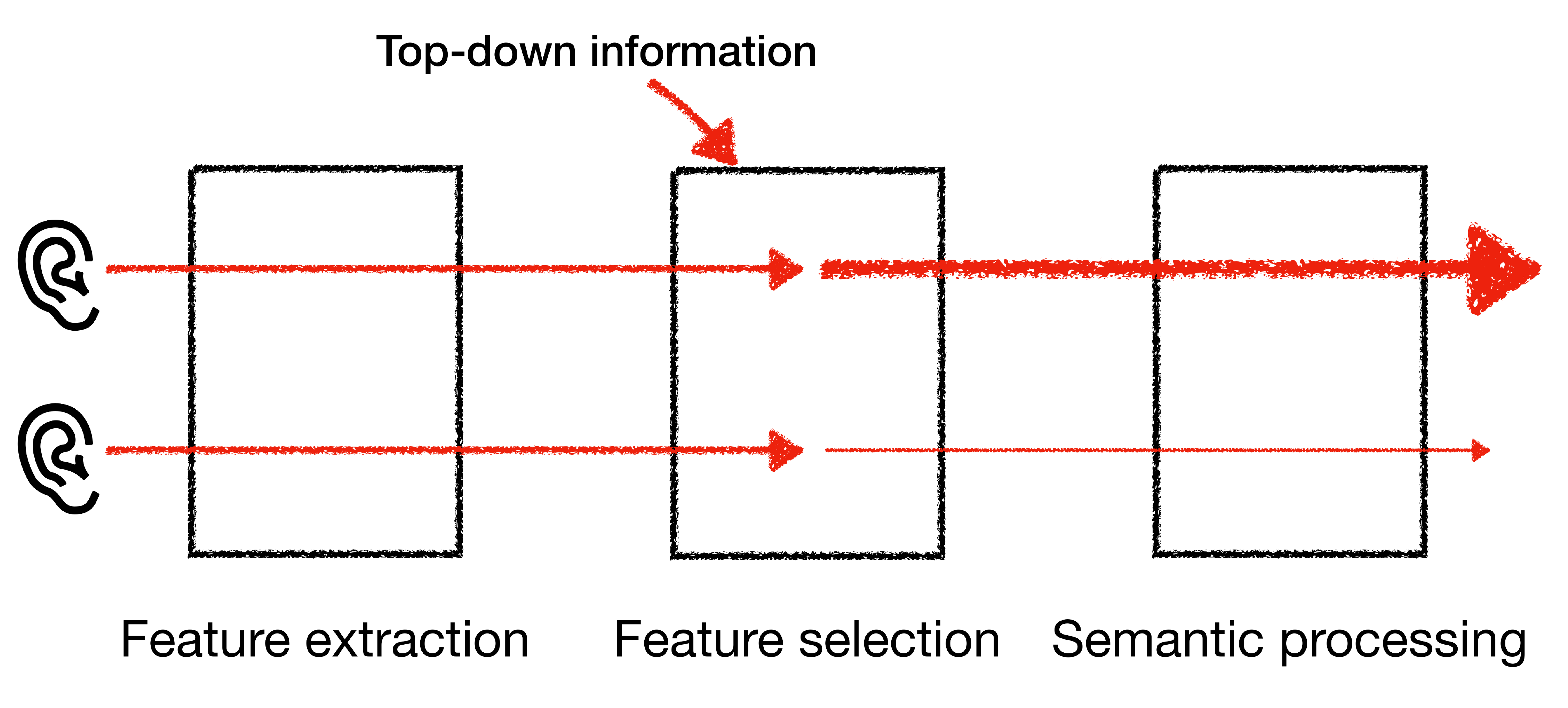}
  \caption{Schematic diagram of selective attention.}
  \label{fig:selective}
\end{figure}

Although these schemes have been validated empirically, the theoretical understanding of their efficiencies is not clear. On the other hand, a theoretical model based on human perception called \emph{selective attention} was proposed by Broadbent \cite{broadbent1958perception} in the 1950s, which leads to a line of research to explain the cocktail party phenomenon \cite{treisman1960contextual,deutsch1963attention,schneider1977controlled}. In its simplest form, this theory suggests that there exists at least one filter in the system for feature selection; bottom-up construction of the feature channels occurs in parallel up to the level of the filter; and beyond the filter, information processing is continued only for those selected channels by the filter. 
The schematic diagram of selective attention is shown in Figure \ref{fig:selective}. Indeed, this selective attention model is supported by neural data~\cite{mesgarani2012selective,gazzaley2012top,akram2016robust}. For example,
when a subject listens to a mixture of two speech signals, the speech spectrogram reconstructed based on neural responses from the auditory cortex reveals only the salient spectral and temporal features of the attended speaker, as if subjects were listening to that speaker alone \cite{mesgarani2012selective}.

Inspired by the selective attention theory and the neuroscience experiments, we propose two deep neural network architectures for speaker extraction based the attention mechanism used by sequence-to-sequence translation \cite{luong2015effective}, which is the foundation of the Transformer network \cite{vaswani2017attention}. Feature or activation selection is achieved by conditioning on the speaker embedding, a form of top-down information, which provides clues on which hidden activations are relevant to the target speaker. 
We name the proposed model architectures AvaTr V1 and AvaTr V2 to emphasize that the speaker embedding is indeed an \emph{avatar} of the target speaker. Both architectures can be understood as performing sequence-to-sequence translation, where the raw waveform of the mixed audio is first converted into a two-dimensional feature representation using a single convolution; the feature representation is then improved and modulated by the speaker embedding and a series of attention-based transformations; and finally, clean speech is predicted by a single deconvolution of the filtered feature representation.   
The differences between these two architectures are where and how we incorporate the speaker embedding. In AvaTr V1, the feature selection is performed once at the early stage. Therefore, we can only take the encoder part of the Transformer, which yields a simple yet efficient model. In AvaTr V2, the architecture is closer to the original Transformer that both the encoder and the decoder are inherited, while the speaker embedding is treated as a query for all attention layers in the decoder, filtering out irrelevant information as we build the feature representation.

The rest of the paper is organized as follows. In section \ref{sec:method}, we present the intuitions and the details of the proposed model architectures. In section \ref{sec:exp}, we validate the proposed model architectures on the LibriSpeech dataset \cite{panayotov2015librispeech}, and specify the experimental settings including the closed-set and open-set regimes.

\section{Our method} \label{sec:method}

The speaker extraction problem is a sequence-to-sequence problem, with mixed audio as an input sequence and separated clean speech of the target speaker as the output sequence
\cite{wang2018voicefilter,xu2020spex,gfeller2020one}. 
A recent family of sequence-to-sequence models called \emph{Transformer} \cite{vaswani2017attention} has shown to be extremely effective in modeling text sequences \cite{brown2020language}, image sequences \cite{dosovitskiy2020image}, and speech separation \cite{chen2020dual}. 
Our proposed models are based on the Transformer architecture, with the main contribution being the implementation of \emph{selective attention} with top-down modulation applied to speaker extraction.

\subsection{Preliminary: Transformer}
The core component in a Transformer is the \emph{attention mechanism} \cite{vaswani2017attention}. It takes as inputs three matrices, $Q \in \mathbb{R}^{N \times d}$, $K \in \mathbb{R}^{M \times d}$ and $V \in \mathbb{R}^{M \times d}$, which represent the sequences of queries, keys and values respectively with $N, M$ the sequence lengths and $d$ the number of feature channels. In general, a \emph{value} $V_i \in \mathbb{R}^d$ and its corresponding \emph{key} $K_i \in \mathbb{R}^d$ are linear projections of the same activation vector at position $i$. In the case of \emph{self-attention}, $Q$, $K$ and $V$ are all computed from the same activation matrix with different linear projections. However, the queries may come from another source making it a good place to insert top-down information.
The output of the attention mechanism is given by   
\begin{align*}
    \text{Attention}(Q, K, V) = W V \,\, \text{with} \,\, W = \text{Softmax}\Big( \frac{Q K^\top}{\sqrt{d}} \Big). 
\end{align*}
Intuitively, the attention weights $W \in \mathbb{R}^{N \times M}$ measure the similarity between any query-key pair, making each row of the output matrix $WV$ a convex combination of the values. Note that the output matrix $WV$ has the same dimensionality as $Q$. Thus, the attention mechanism may also be interpreted as a ``transformation'' of the queries. 

The Transformer architecture usually consists of an \emph{encoder} and a \emph{decoder}. The encoder is made of $L$ \emph{self-attention blocks} of the form:
\vspace{1mm}
\begin{eqblck}{{Self-Attention Block: $Z^l \mapsto Z^{l+1}$}}
\vspace{-3mm}
\begin{align*}
    &Q, K, V = \text{Linear}(Z^l), \, \text{Linear}(Z^l), \,  \text{Linear}(Z^l) \\
    &\widehat{Z}^{l} = \text{LayerNorm}\Big( Z^l + \text{Dropout}\big( \text{Attention}(Q, K, V) \big) \Big) \\
    &Z^{l+1} = \text{LayerNorm}\Big( \widehat{Z}^l + \text{Dropout}\big( \text{MLP}(\widehat{Z}^l) \big) \Big)
\end{align*}
\end{eqblck}
Conditioned on the self-attention output $Z^L$ of the encoder, the decoder is made of a series of \emph{conditional attention blocks} of the form:
\begin{eqblck}{{Conditional Attention Block: $Y^l, Z^L \mapsto Y^{l+1}$}}
\vspace{-3mm}
\begin{align*}
    &Q, K, V = \text{Linear}(Y^l), \, \text{Linear}(Y^l), \, \text{Linear}(Y^l) \\
    &\widetilde{Y}^{l} = \text{LayerNorm}\Big( Y^l + \text{Dropout}\big( \text{Attention}(Q, K, V) \big) \Big) \\
    &\widetilde{Q}, \widetilde{K}, \widetilde{V} = \text{Linear}(\widetilde{Y}^l), \, \text{Linear}(Z^L), \, \text{Linear}(Z^L) \\
    &\widehat{Y}^{l} = \text{LayerNorm}\Big( \widetilde{Y}^l + \text{Dropout}\big( \text{Attention}(\widetilde{Q}, \widetilde{K}, \widetilde{V}) \big) \Big) \\
    &Y^{l+1} = \text{LayerNorm}\Big( \widehat{Y}^l + \text{Dropout}\big( \text{MLP}(\widehat{Y}^l) \big) \Big)
\end{align*}
\end{eqblck}


Other common practices include adding \emph{positional encoding} before applying attention operations, and extending the vanilla attention to \emph{multi-head attention} (MHA), which involves parallel attentions on disjoint groups of the feature channels (a.k.a. heads), followed by a joint operation to fuse all groups.


\subsection{Overview}

\begin{figure*}[t]
    \centering
    \begin{minipage}[b]{0.30\linewidth}
    \centering
    \includegraphics[width=\textwidth]{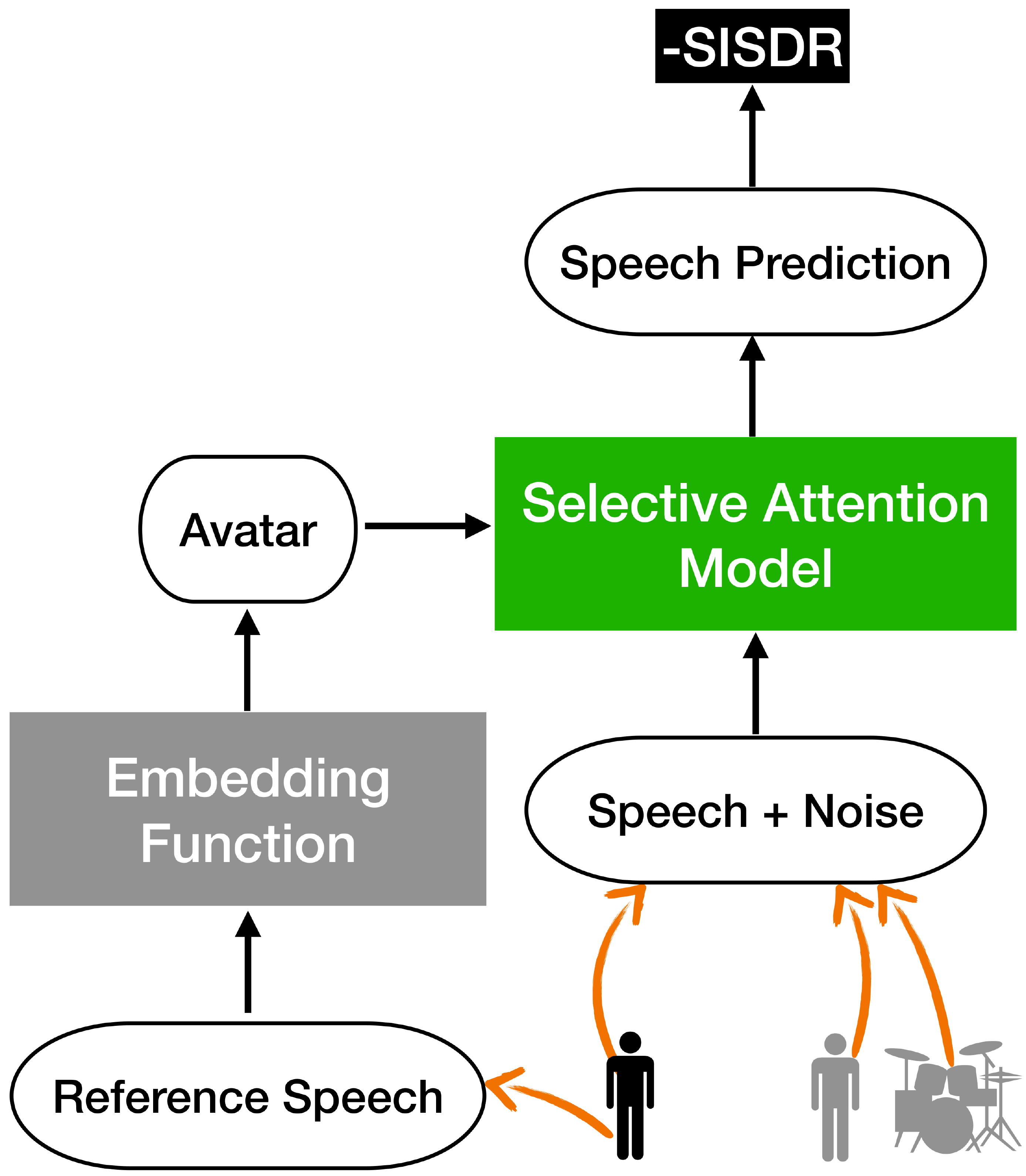}
    (a) Overview
    \end{minipage}
    \hspace{0.5cm}
    \begin{minipage}[b]{0.30\linewidth}
    \centering
    \includegraphics[width=\textwidth]{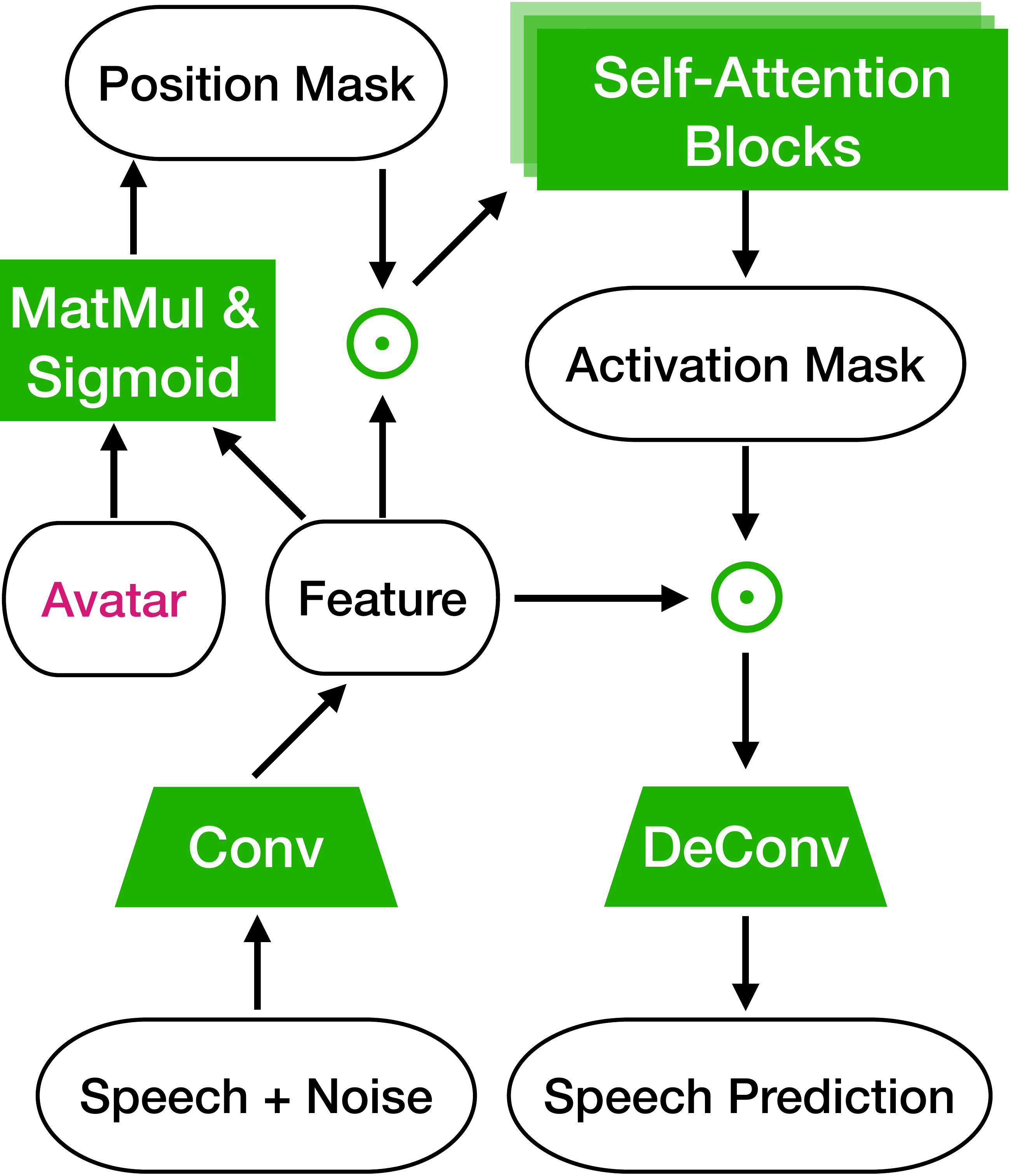}
    (b) AvaTr V1
    \end{minipage}
    \hspace{0.5cm}
    \begin{minipage}[b]{0.30\linewidth}
    \centering
    \includegraphics[width=\textwidth]{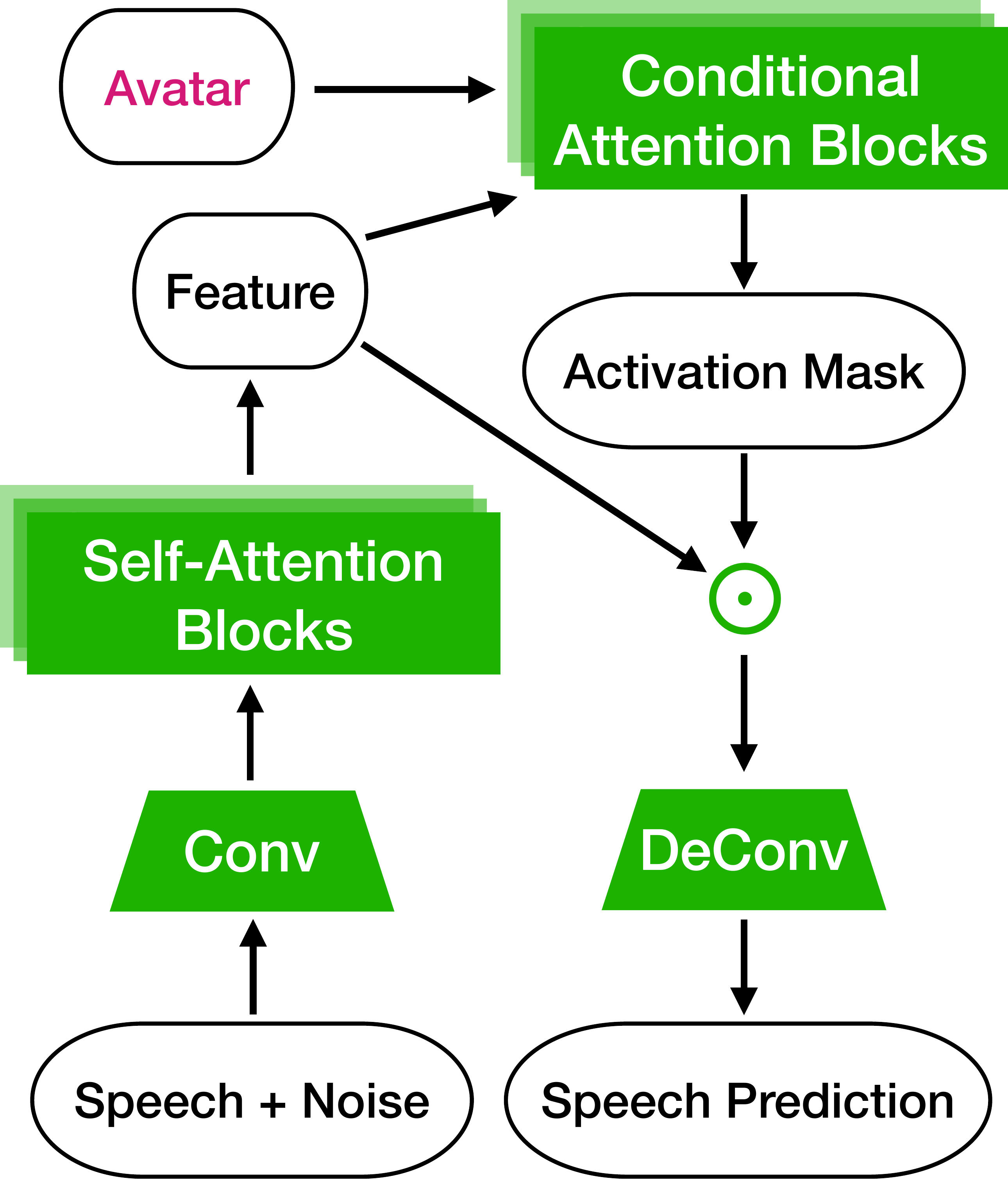}
    (c) AvaTr V2
    \end{minipage}
    \caption{Our proposed model architectures, which are named as AvaTr V1 \& V2 respectively. 
    (a) gives an overview of the speaker extraction system. (b) and (c) are two instances of the selective attention model in (a).}
    \label{fig:avatr}
\end{figure*}

We start by describing a schematic pipeline for implementing the selective attention theory with speaker induced top-down modulation, shown in Figure \ref{fig:avatr}(a). We assume that a mixed audio is generated by mixing a \textbf{clean speech} of a target speaker and various \textbf{noise} including interfering speech from other speakers and ambient noise. To extract clean speech from the mixed audio, we assume that a very short \textbf{reference speech} (thus is called one-shot) from the target speaker is also provided, which we feed to the \emph{embedding function} to obtain an \emph{avatar} (i.e., a speaker embedding) of the target speaker. Specifically, the avatar is a vector, summarizing the characteristics of the speaker depending on the choice of the embedding function. In this paper, we instantiate the embedding function using the Wav2Vec2 model \cite{baevski2020wav2vec}.

The main module of our speaker extraction framework is the \emph{selective attention model}, which takes as inputs the mixed audio and the avatar to predict the clean speech. The avatar serves as a top-down information to modulate the feature representation of the mixed audio, which amounts to softly mask out activations that do not belong to the target speaker. We propose in the next subsections two instances of the selective attention model, which are named AvaTr V1 and V2.

To summarize, there are two learnable modules in the proposed framework: the embedding function and the selective attention model. Our loss function is the negative \emph{scale-invariant signal-to-distortion} (SISDR) \cite{le2019sdr}. 
We use a single convolution and correspondingly a single ``deconvolution'' (i.e., convolution transpose) to map between the audio space and the feature space.

\subsection{AvaTr V1}

The idea of AvaTr V1 is to build up the activation mask progressively. The \emph{Avatar} is a vector in $\mathbb{R}^d$ and the feature representation comes from an initial convolutional layer, denoted by \emph{Feature} in $\mathbb{R}^{N \times d}$.
We compute a \emph{PositionMask} and use it to modulate \emph{Feature} by 
\begin{align}
    &\text{PositionMask} = \text{Sigmoid}(\text{Feature} \cdot  \text{Avatar}), \\
    &\overline{\text{Feature}} = \text{Feature} \odot \big( \text{PositionMask} \cdot \mathbf{1}_d^\top \big),
\end{align}
This is then refined by a series of self-attention blocks, denoted by $\text{SAB}_1, \ldots, \text{SAB}_L$, to generate the \emph{ActivationMask}: 
\begin{align}
    \text{ActivationMask} = \text{SAB}_L \circ \cdots \circ \text{SAB}_1\big( \overline{\text{Feature}} \big).
\end{align}
Applying the activation mask on the initial feature representation we obtain the final selected features. This gives a rough estimate of how likely an activation is caused by the target speaker. The architecture of AvaTr V1 is shown in Figure~\ref{fig:avatr}(b). 

Note that this approach uses the avatar in an early stage. This is in line with Broadbent's theory \cite{broadbent1958perception}, which argues that feature selection occurs before semantic information processing: it creates an information bottleneck to prevent the system from becoming overloaded. 

\subsection{AvaTr V2}
One potential issue of AvaTr V1 is that the top-down modulation is executed only once in the early stage of the network, which may not take full advantage of the top-down information. 
As also pointed out by Treisman et al. \cite{treisman1960contextual}, Broadbent's theory cannot fully account for the cocktail party effect, as psychological experiments show that participants can still process the meaning of the unattended messages.  

To circumvent these issues, we modify AvaTr V1 with two improvements. 1) We insert a few self-attention blocks after the initial convolution and before the top-down modulation. This incorporates contextual information into the feature representation of the mixed audio, leading to a more precise similarity measurement with respect to the avatar. 
2) We add a series of modified conditional attention blocks, denoted by $\text{CAB}_1, \ldots, \text{CAB}_K$, which takes as inputs the feature representation constructed using the self-attention blocks and additionally the avatar, then outputs the activation mask as
    \begin{align}
        &\text{ActivationMask} = \text{CAB}_K( Y^{K-1}, \text{Feature}, \text{Avatar} ) \\
        &\quad \text{with}\,\, Y^{k} = \text{CAB}_k( Y^{k-1}, \text{Feature}, \text{Avatar} ) \,\,\text{and}\,\, Y^0 = \mathbf{0}, \notag
    \end{align}
where $\text{Feature}$ is the final output of the self-attention blocks.
To embed the avatar into the conditional attention block, we modify the queries of the first attention layer of each CAB as 
\begin{align}
    Q = \text{Linear}\big( Y^k + \mathbf{1}_N \cdot \text{Avatar}^\top \big).
\end{align}
The avatar is therefore used multiple times as we generate the activation mask. The architecture of AvaTr V2 is depicted in Figure \ref{fig:avatr}(c). 

\subsection{Model training}

Unlike speech separation, which requires a permutation invariant training \cite{yu2017permutation}, the model training for speaker extraction can be simply supervised as we only look for the signal defined by the avatar of the target speaker. Specifically, we simulate mixed audios from clean speech and interfering noises, where the clean speech provide the ground truth, which is then used to compute the negative SISDR with respect to the output of the selective attention model.


The speaker extraction problem proposed in this paper is similar to a few-shot learning problem \cite{vinyals2016matching} in that each speaker defines a unique task and the goal is to learn a model that can generalize to new speakers when one shot of the clean speech is available. To this end, we formulate the model training to be episodic, with input episodes of the form 
$
\big(\text{Mixed audio}, \text{reference speech} \big), 
$
which serves as the ``query'' set and the ``support'' set respectively in analogy to the terminology of few-shot learning. In training, we are also provided the ground truth of the simulated ``query'' set. The episodic training amounts to minimizing 
\begin{align}
    \mathbb{E}_{S} \mathbb{E}_{N} \mathbb{E}_{R} \Big[ -\text{SISDR}\big( S, \text{Model}(S + N, R) \big) \Big],
\end{align}
where $S, N, R$ represent clean speech, noise and reference respectively. 

In general, the sets of speakers for generating the episodes are disjoint for training and testing. We name this setting the \emph{open-set} regime. However, we may also consider the case where the set of speakers remains the same for both training and testing. We call this setting the \emph{closed-set} regime. Note that these two regimes only differ in evaluation.

\section{Experiments} \label{sec:exp}

\begin{table*}[ht]
  \begin{minipage}{.65\linewidth}
  \centering
  \begin{tabular}{l|l|cccc}
    \toprule
     &  & VF \cite{wang2018voicefilter} & SF \cite{gfeller2020one} & AvaTr V1 & AvaTr V2 \\
    \hline
    \multirow{2}{*}{S + N} & Open & $11.02 \pm 0.01$ & $13.10 \pm 0.02$ & $14.48 \pm 0.01$ & $\textbf{15.02} \pm 0.01$
    \\ & Closed & $10.54 \pm 0.002$ & $13.43 \pm 0.004$ & $14.76 \pm 0.002$ & $\textbf{14.98} \pm 0.004$ \\
    \hline
    \multirow{2}{*}{S + S} & Open & $8.08 \pm 0.04$ & $13.89 \pm 0.09$ & $16.64 \pm 0.07$ & $\textbf{17.44} \pm 0.14$
    \\ & Closed & $8.49 \pm 0.04$ & $14.90 \pm 0.02$ & $16.27 \pm 0.07$ & $\textbf{18.38} \pm 0.06$ \\
    \hline
    \multirow{2}{*}{S + A} & Open & $6.72 \pm 0.03$ & $10.66 \pm 0.07$ & $12.94 \pm 0.08$ & $\textbf{13.52} \pm 0.09$
    \\ & Closed & $7.24 \pm 0.02$ & $11.67 \pm 0.01$ & $13.52 \pm 0.03$ & $\textbf{13.76} \pm 0.03$ \\
    \bottomrule
  \end{tabular}
  \caption{Results (SISDR) on LibriSpeech. The baselines are VoiceFilter (VF) and SoundFilter (SF). For AvaTr V1, we use $5$ SABs and hidden size $256$. We try 3 types of mixed audios: speech vs. noise (S+N), speech vs. interfering speech (S+S) and speech vs. all (S+A), and evaluate all models with both the open-set and closed-set regimes. 
  }
  \label{tab:librispeech}
  \end{minipage} 
  \hfill
  \begin{minipage}{.32\linewidth}
  \centering
  \begin{tabular}{l|l|c}
    \toprule
     &  & AvaTr V1 \\
    \hline
    \multirow{2}{*}{\# Blocks} & 1 & $10.38 \pm 0.06$ \\ & 3 & $15.12 \pm 0.04$ \\ & 5 & $16.64 \pm 0.07$ \\ & 8 & $\textbf{17.64} \pm 0.09$ \\
    \hline
    \multirow{2}{*}{Hidden Size} & 128 & $13.93 \pm 0.06$ \\ & 256 & $16.64 \pm 0.07$ \\ & 512 & $\textbf{16.74} \pm 0.07$ \\
    \bottomrule
  \end{tabular}
  \caption{Ablation study for AvaTr V1 on mixture speech vs. speech with different numbers of SABs and hidden sizes.}
  \label{tab:ablation}
  \end{minipage}
\end{table*}


\subsection{Setup}

Since it is hard to collect both the mixed audio and the ground-truth clean speech, we simulate mixed audios with randomly sampled clean speech and noises. We consider three mixture types: a) speech vs. interfering speech (S + S), b) speech vs. ambient noise (S + N), and c) speech vs. interfering speech + ambient noise (S + A). 

We first describe the open-set regime, where a model is tested in its generalization to unseen speakers and novel noise categories.  
For training, we sample clean speech and interfering speech from the train-clean-100 subset of the LibriSpeech dataset \cite{panayotov2015librispeech} 
and we sample ambient noise from the AudioSet \cite{gemmeke2017audio}. This involves 251 speakers and 518 non-speech classes.   
For validation and testing, all speech is sampled from the dev-clean (40 speakers) and test-clean (40 speakers) subsets respectively, while the ambient noise is sampled from the union of the WHAM dataset \cite{wichern2019wham} and the ESC50 dataset \cite{piczak2015esc}. 
Note that these tested speakers do not overlap with training speakers. 

We also consider the closed-set regime, where the setup is exactly the same as that of the open-set regime except the set of testing speakers remains the same as the training speakers. It serves as a sanity check to see how well a model generalizes to unseen speech of the same speakers. To this end, we randomly select 3000 clean speech episodes each from train-clean-100 subset as the validation set, and 3000 other samples from the same speakers as the test set. The interfering speech and ambient noises are drawn exactly as in the open-set regime. 

In order to take full advantage of the mixture simulation, all models are trained with online mixing. Following the setup of \cite{gfeller2020one}, the mixed audios are generated at varying SNRs, which are chosen uniformly at random from $[-4, 4]$ dB. 
For validation and testing, we generate mixtures with SNR of $0$ dB to remove randomness in evaluation. 
We train our models with mixed audio of $3$ second duration, and evaluate the model performance on full-length audio. For computing the avatar, we sample $2$-second reference speech from the same audiobook where the clean speech is sampled from.



\subsection{Results}


Our open-set and closed-set results are presented in Table \ref{tab:librispeech}. We compare the proposed AvaTr V1 and V2 with two state-of-the-art methods, VoiceFilter \cite{wang2018voicefilter} and SoundFilter \cite{gfeller2020one}. Since neither VoiceFilter nor SoundFilter released code, we tried our best to reproduce their results. 

\begin{figure}[h]
    \centering
    \includegraphics[width=0.9\linewidth]{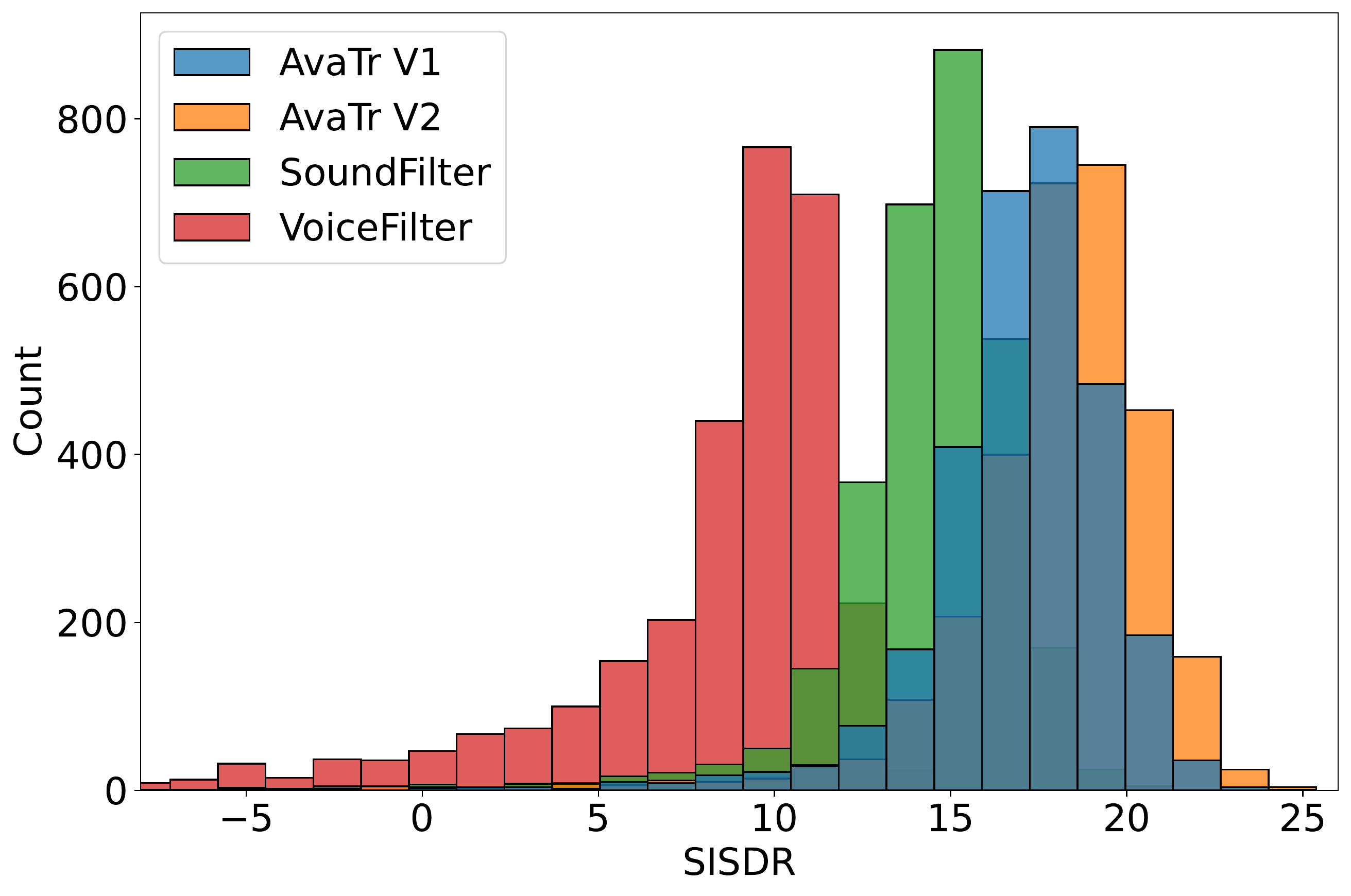}
    \caption{Histograms of SISDR on speaker extraction (speech vs. speech) for over many examples of mixed audio, for our networks and state-of-the-art competitors.}
    \label{fig:hist}
\end{figure}

Figure~\ref{fig:hist} shows histograms of SISDR over multiple speech separation trials for all methods (for speech vs. speech mixtures), and Table \ref{tab:librispeech} shows overall performance. It can be seen that both AvaTr V1 and V2 outperform the state-of-the-art methods, which shows that Transformer-based architectures have great potential in speaker extraction. 

\begin{figure}[h]
    \centering
    \includegraphics[width=0.9\linewidth]{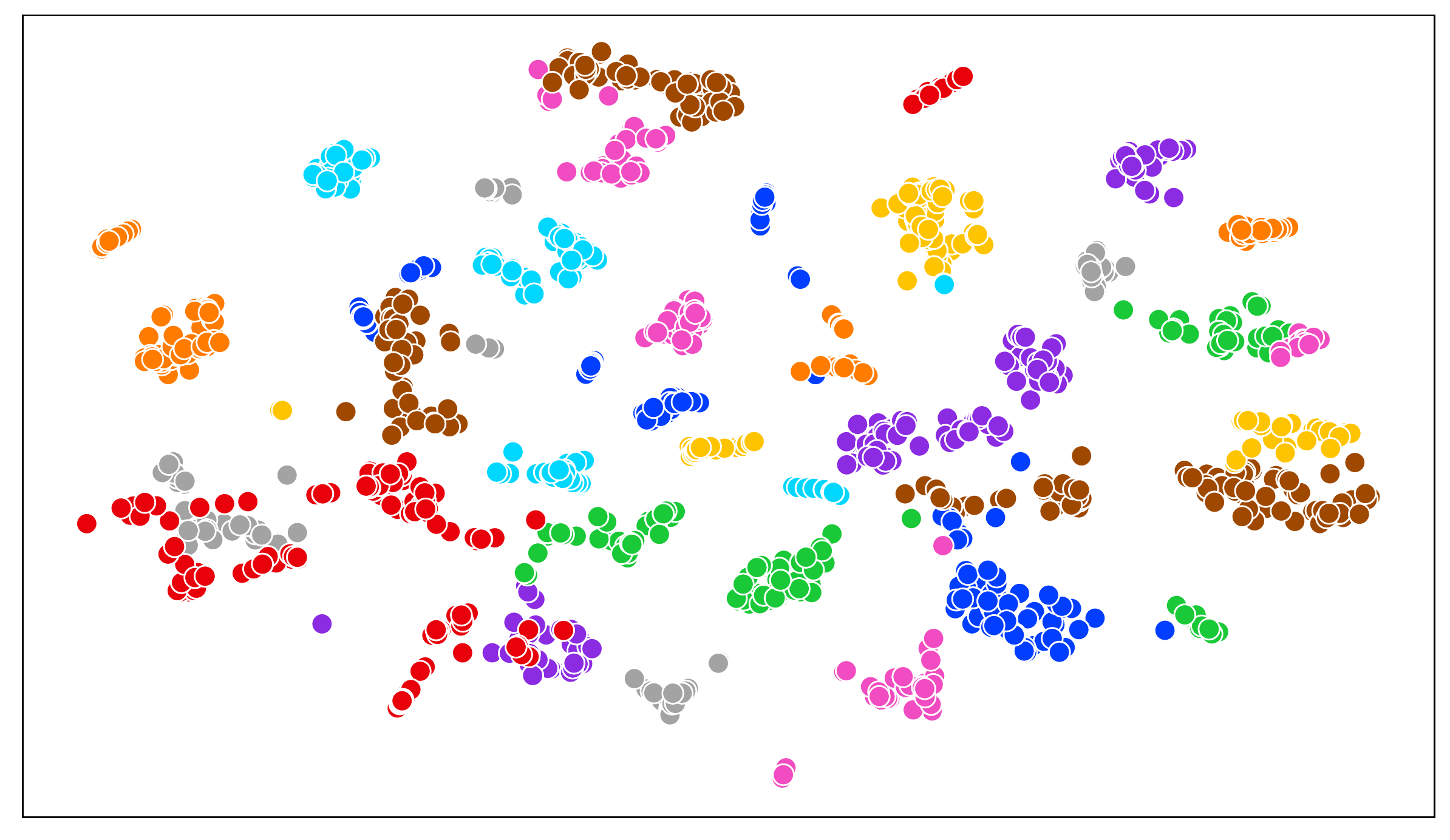}
    \caption{t-SNE plot of learned avatars from 40 speakers colored by speaker ID.}
    \label{fig:t-SNE}
\end{figure}


In addition, we plot the avatars learned by AvaTr V1 for speech vs. speech in Figure~\ref{fig:t-SNE}, and provide audio samples in the supplemental material for a qualitative comparison.

\subsection{Ablation study}
To understand the impact of design choices of different architectural hyper-parameters like number of attention blocks and hidden layer size, we did some ablation study using AvaTr V1 for the case of speech vs. speech. The result is shown in Table~\ref{tab:ablation}. 

\section{Conclusions}
We proposed a framework for speaker extraction inspired by the theory of selective attention. Based on the Transformer network, a conditional attention mechanism is used to incorporate top-down speaker information obtained from a brief duration of clean reference speech. 
We demonstrated the efficiency of our models on the LibriSpeech dataset.


\bibliographystyle{IEEEtran}

\bibliography{mybib}

\newpage

\section{Supplemental Material}

\subsection{Experimental Details}
We implemented our models using  PyTorch~\cite{paszke2017automatic} and trained on 4 NVIDIA RTX 3090 GPUs with memory of 24GB each. The models are trained for 500 epochs with a learning rate scheduler which reduces the learning rate by half when validation loss plateaus. We use ADAM~\cite{kingma2014adam} optimizer with default parameters and initial learning rate of $10^{-4}$ and batch size of $16$. The batch size chosen experimentally as shown in Table~\ref{tab:batch_size_vs_sisnr} using Speech vs Speech mixtures on LibriSpeech dataset.

\begin{table}[ht]
  \centering
  \begin{tabular}{cc}
    \toprule
    Batch Size & SISNR \\
    \hline
    4 & 16.44$\pm$ 0.21 \\
    16 & \textbf{16.64}$\pm$ 0.07 \\
    32 & 16.08 $\pm$ 0.04 \\
    \bottomrule
  \end{tabular}
  \caption{AvaTr V1 Performance in \emph{SISDR} while varying batch sizes on LibriSpeech dataset}
  \label{tab:batch_size_vs_sisnr}
\end{table}

\subsection{Extra Results on LibriSpeech}
\begin{figure}[h]
    \centering
    \includegraphics[width=0.9\linewidth]{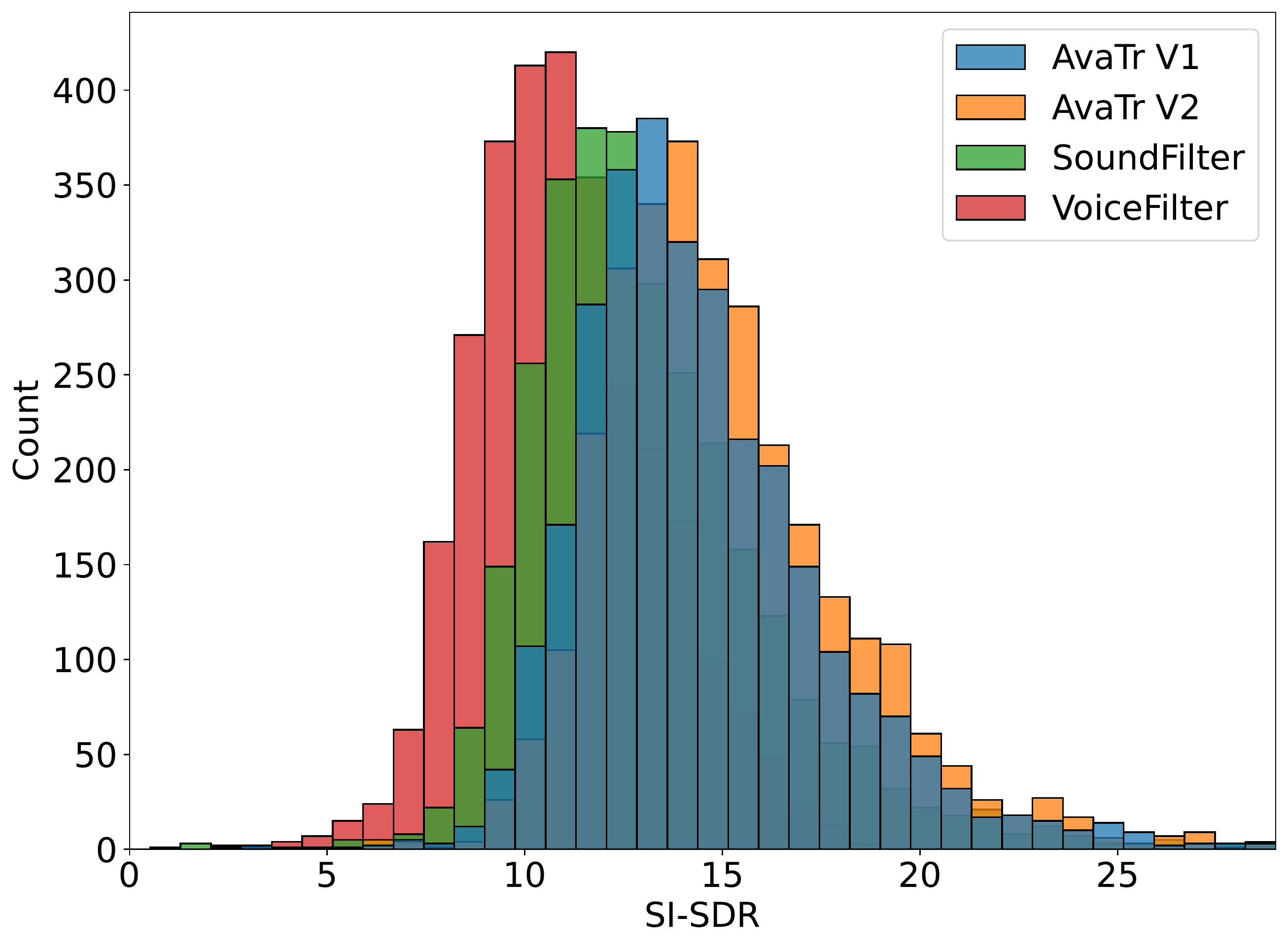}
    \caption{Histograms of SISDR - Speech vs Noise}
    \label{fig:hist_vs_noise}
\end{figure}

\begin{figure}[h]
    \centering
    \includegraphics[width=0.9\linewidth]{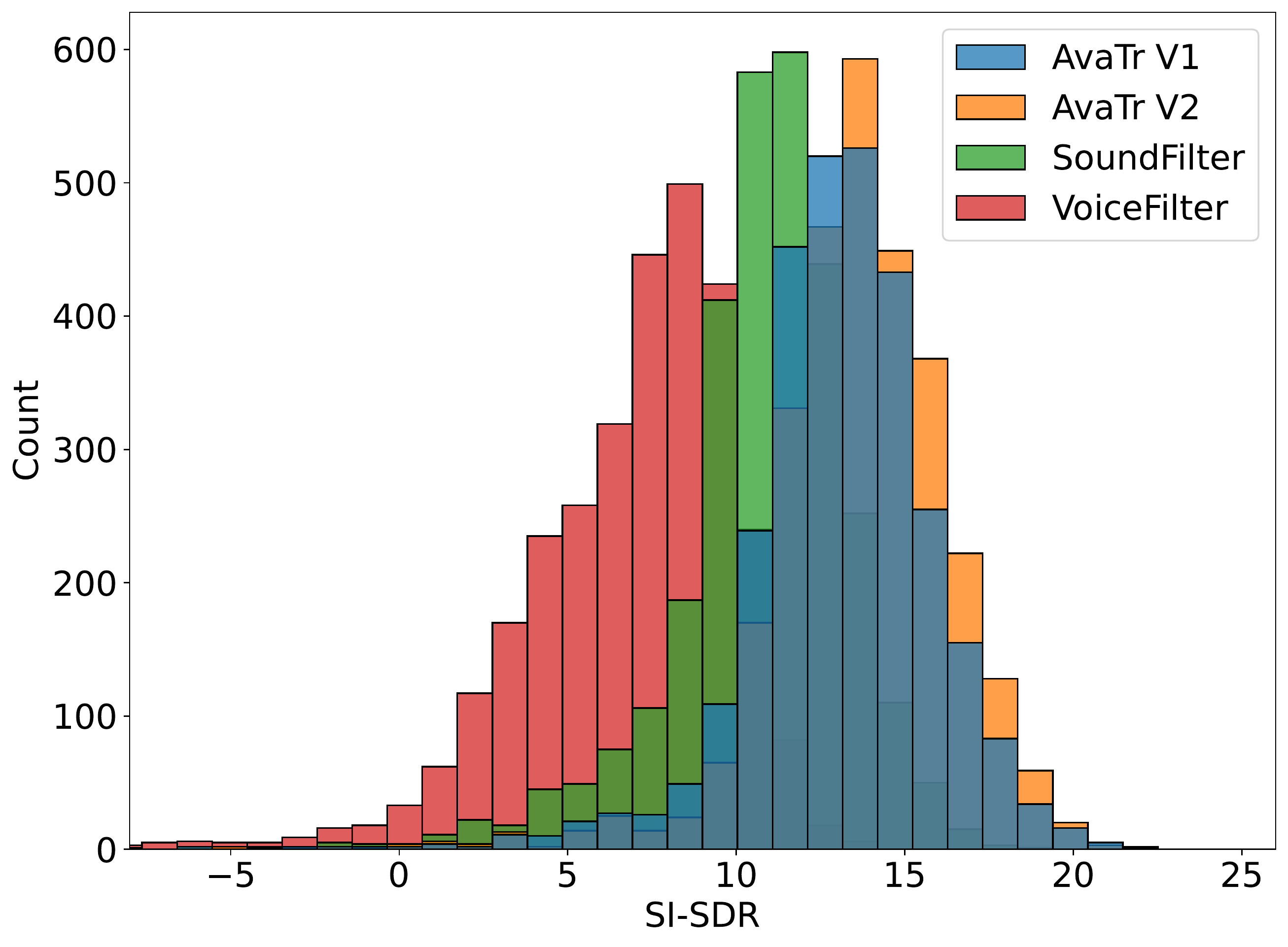}
    \caption{Histograms of SISDR - Speech vs All (interfering speech+noise)}
    \label{fig:hist_vs_all}
\end{figure}

This section highlights the results of histograms (Figure~\ref{fig:hist_vs_noise} and ~\ref{fig:hist_vs_all}) of SISDR over all methods (for speech vs. noises and speech vs. all mixtures).The results indicate that the variability in the case of AvaTr V1 and AvaTr V2 is much smaller than other methods thus proving their superiority. Figure~\ref{fig:hist_vs_all} also indicate that selective attention conditioned on avatar provides much more stable performance than others when we have an interfering speech signal with noises.

\subsection{Results on FSD50K}
Since we are trying to address the problem of signal extraction using another reference signal, all of the methods we studied in our work should also be able to extract other sounds than human speech. For this reason we also trained models on FSD-50K dataset \cite{eduardo_fonseca_2020_4060432} which contains environmental recordings of sound events drawn from the AudioSet Ontology. The main experimental set-up is similar to what we have done with LibriSpeech. However, due to the great diversity of natural sounds and FSD-50K has multi-class sound
annotations, we extract 2 seconds of audio for reference from the same audio example with target sound. The obtained SISDR scores are showed in Table~\ref{tab:FSD}.

\begin{table}[ht]
  \centering
  \begin{tabular}{l|cccc}
    \toprule
     & VF  & SF  & AvaTr V1 & AvaTr V2 \\
    \hline
    FSD-50K & $6.13$ & $8.76$ & $\textbf{8.97}$ & $8.83$
    \\
    \bottomrule
  \end{tabular}
  \caption{Results (SISDR) on FSD-50K dataset.
  }
  \label{tab:FSD}
\end{table}

\end{document}